\documentclass{jfm}
\usepackage{float}
\usepackage{graphicx}
\usepackage{epstopdf,epsfig}
\usepackage{newtxtext}
\usepackage{newtxmath}
\usepackage{natbib}
\usepackage{hyperref}
\hypersetup{
   colorlinks = true,
   urlcolor   = blue,
   citecolor  = black,
}

\newcommand{\RomanNumeralCaps}[1]

\title{Dynamics of a film flowing down a granular chain}

\author{Kishorkumar Sarva\aff{1}\corresp{\email{kishorsarva@iisc.ac.in}}}

\affiliation{\aff{1} Interdisciplinary Center for Energy Research (ICER), IISc Bangalore, 560012, India 
 
 }
\begin{document}
\maketitle

\begin{abstract}
We investigate the effects of fibre morphologies, such as single and granular chain with torus bead on liquid film evolution using experimental and axi-symmetric numerical simulations with a one-fluid formulation. We introduce a non-dimensional parameter 'Bead Ratio'($BR$), that is, the ratio of bead diameter to the film height. When both the $BR$ and its distance from the nozzle exceed critical values,  selection mechanism leading to development of 'dominating' waves: from regularly spaced droplets to coarsening or droplet merging. The two mechanisms, influenced by the $BR$ for a single bead, contribute to droplet merging: the formation of the downstream healing length, which precipitates the initial stage, and it's oscillating behaviour resulting in droplet merging. When the bead position is away from the healing length far from the nozzle, the transient simulations capture behavior similar to the finite amplitude perturbations at the inlet. However, when the bead is within the healing length, the film evolution has only coarsening effect on the droplet spacing. When the bead spacing on a granular chain is less than the droplet spacing of a Rayleigh-Plateau regime is significantly altered.

\end{abstract}


{\bf MSC Codes }  {\it(Optional)} Please enter your MSC Codes here

\section{Introduction}
A fascinating phenomenon that occurs in a variety of industrial processes and natural systems is gravity-driven liquid film flow on cylindrical fibres. These films show complex interfacial dynamics. The intricate dynamics of this flow behaviour, especially at low Reynolds numbers ($Re$), are essential for comprehending the open-flow issues covered in earlier studies [\cite{duprat2007absolute}]. Given that both are caused by the Rayleigh-Plateau instability, \citep{gallaire2017fluid} it is closely related to the liquid jet problem. Ink-jet printing, biomedical procedures, and falling film formation in coatings for desalination and gas absorption (\cite{Grunig2012mass}) highlight the importance of understanding film formation and its dynamics. Droplets form on the surface of the fibre in the current issue as opposed to liquid jets, where the breakup of the jet causes droplets to form. A number of control parameters affect the properties of the film, including droplet size, spacing, and film thickness. \\
\begin{figure} 
   \centering
   \captionsetup{width=\linewidth}
  \captionsetup{justification=justified} 
  \includegraphics[width=0.8\textwidth]{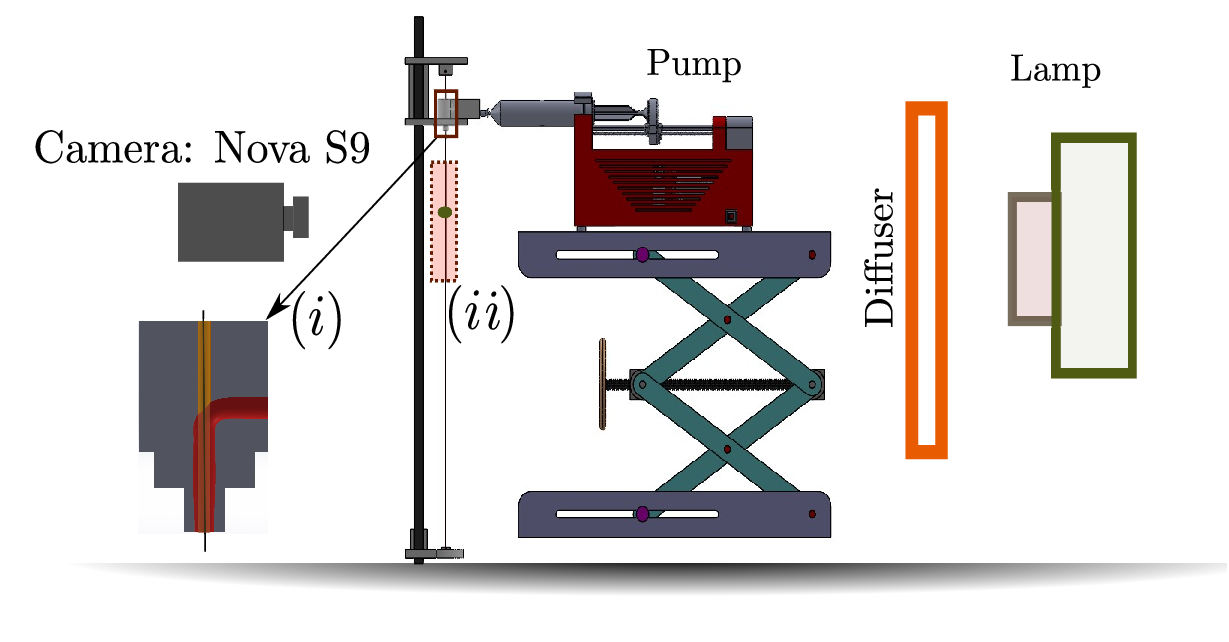}
  \caption{The experimental setup for high-speed imaging is depicted. The various configuration modules are labeled in the diagram. The co-axial nozzle indicated with ($i$), generates a coaxial flow field on the fibre. The yellow rectangle with shade on the fibre with bead represents with $(ii)$ is the region of interest.}
  \label{fig:testrig} 
\end{figure}

Key insights into influence of control parameters such as fluid proprieties, flow rate, inlet perturbation, nozzle to fibre diameter ratio have gained over the years. The film formation for highly viscous fluids have been gained from experimental observations by  \cite{kliakhandler2001viscous}.  The study showed that the film will exhibit three regimes based on the flow rate conditions namely, bead merging (convective regime), Rayleigh Plateau instability (chain of droplets) and isolated bead. \cite{sisoev2006film} discovered attracting wave regimes and dominant waves in fiber flow, influenced by system parameters and forcing conditions. Inlet perturbations and their impact on flow behavior have been further examined by \cite{duprat2009spatial}.They observed the speed of the travelling waves  decrease with decrease in forcing frequency which is estimated from the average slope of the crest lines on the spatio-temporal diagram. Forcing amplitude also has influence on the characteristics of traveling waves. The healing length decreases with increase in the amplitude, and increases coalescence events near to the nozzle. The influence of the nozzle-to-fiber ratio on regime transitions has been identified ( \cite{ji2020modelling}, where impacts the regime transitions as larger the nozzle diameter convective to Rayleigh-Plateau regime. The study also revealed that healing length decreases with increase in nozzle diameter and increase in the droplet spacing.  Improved modeling approaches involving the Hamacker constant and domain transformation techniques have enhanced agreement with experimental results (\cite{ji2020modelling} ; \cite{liu2021coating}). These studies collectively provide a comprehensive understanding of film flow down a fiber, shedding light on the emergence of wave regimes, dominant wave selection, and the influence of system parameters and forcing conditions.

The studies on the present problem through various methodologies and different operating conditions, has shown that the film dynamics depends on various parameters such as fiber diameter ($D_f$), nozzle diameter ($D_n$) and inlet flow conditions (constant flow rate or perturbed velocity or interface) and fluid properties.  Recent studies related to fiber geometries like elliptical \citep{li2017viscous}, and conical fiber  \citep{chan2021film} and \citep{xie2021investigation} are also found important consequences. The conical fiber shown to have significant effect on droplet formation and the size distribution. The current study investigates the dynamics of liquid films on influence of morphology of cylindrical fibres using both experimental and numerical methodologies and under various operating conditions, elucidating the effects of variables including nozzle diameter ($D_n$) to fibre diameter ($D_f$),  inlet flow conditions (constant flow rate, perturbed velocity). Significant changes in film morphology and critical conditions have been found as a result of recent investigations on fibre geometries by \citep{xie2021investigation}. Instead of observing regime transitions, these studies showed on variations in droplet spacing brought on by changes in fibre morphology. 

\begin{figure} 
  \centering
 \captionsetup{justification=justified} 
\includegraphics[width=0.75\textwidth]{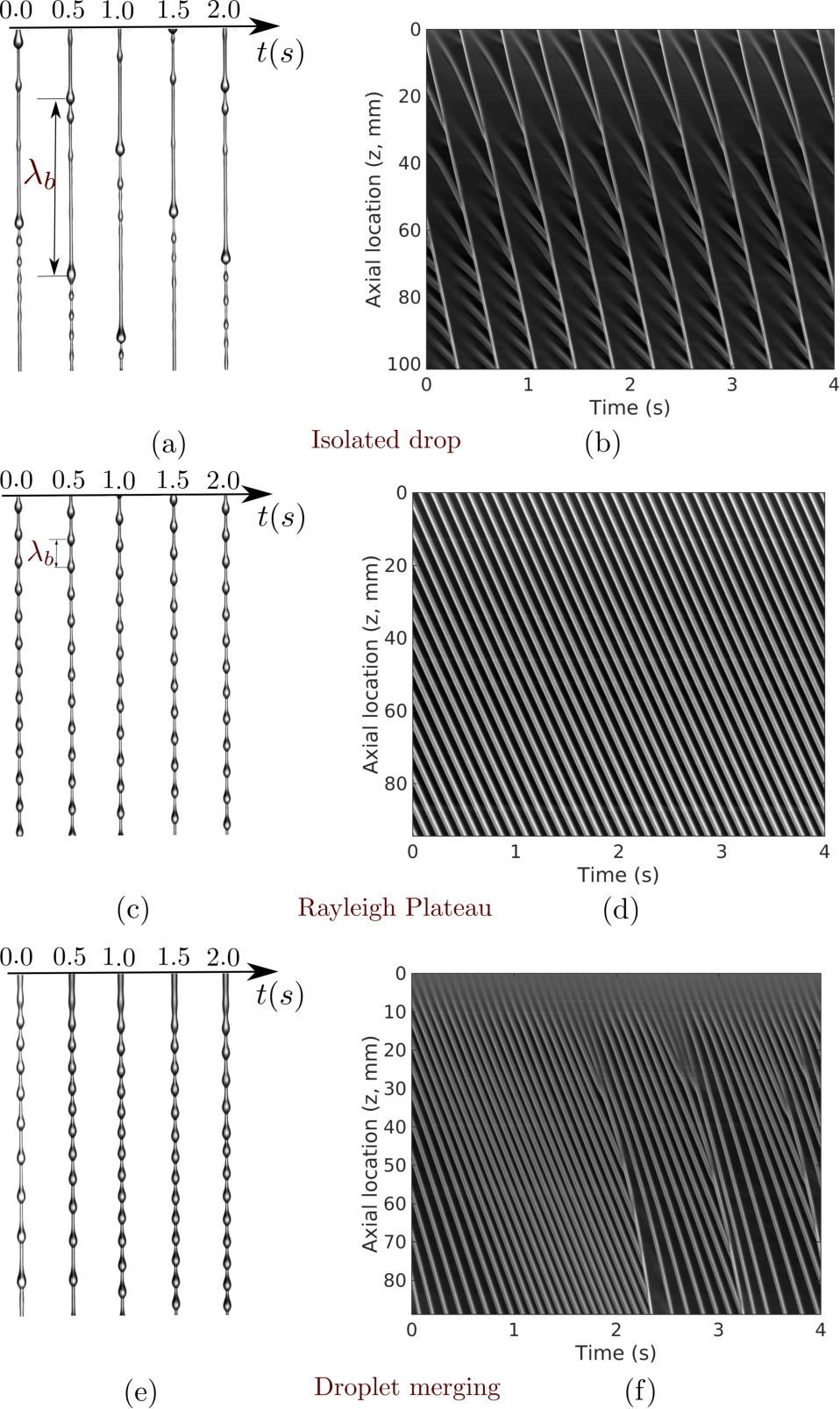}
  \captionsetup{width=\linewidth}
  \caption{ Time sequence images shown for increasing Reynolds number and their respective space-time evolution of the film. (a,b) Isolated droplet with residual film creating ripples at $Re=0.13$. (c,d) Rayleigh Plateau at $Re=0.155$, showing a chain of droplets with equal spacing on the flat cylindrical fiber. (e,f) Droplet merging events at $Re=0.22$, with advecting droplets at different speeds on the fiber. The intrinsic length scale is the fiber diameter of $0.35$ mm. Time stamps are provided in seconds. }  
  \label{fig:flatfibre_events}  
\end{figure}
In our work, we introduce a bead onto the fibre and analyse the resulting dynamics to address the impact of fibre morphology. First, we examine the steady-state behaviour and choice of dominant waves for a liquid film on a flat cylindrical fibre and a fibre with a single bead under identical conditions. Additionally, we compare direct numerical simulations (DNS) with the experimental observations of the liquid film. We continue to investigate the nonlinear evolution for various bead ratios ($BR$) and positions of the bead, which reveals different system final stages depending on the size and location of the bead on the fibre. Finally, we demonstrate the development of the film on a granular fibre.

\section{Experimental setup and observations}

Figure \ref{fig:testrig} shows a schematic illustration of the high-speed imaging setup used for studying variations in the thickness of a film flowing down on a fiber. The various components of the experimental setup are  labelled in the schematic diagram. The main test rig consists a syringe pump (New Era Pump Systems, NE-1000), a syringe and a hypodermic needle, assembly of vertical fibre and tank collecting the dispensed fluid. Shadowgraphy module for visualising the fluid film evolution using high-speed camera (Photron Fastcam Nova SA9) is connected through a computer (i7 processor) to an image acquisition software. A diffuser plate is used in front of an LED source. In the present experimental study, we use Silicone oil with density, $\rho = 960$ kg$/$m$^3$), viscosity $\mu = 48$ mPa$\cdot$S, and surface tension coefficient $\gamma = 20.6$ mN$/$m. The nozzle to fibre diameter ratio is set to $D_n/D_f=4.7 \& 14.28$ which is  similar to the experimental conditions of \cite{ji2020modelling}. 
The parameters that govern the physics of the above system are the liquid properties, the nozzle size $D_n$, fibre diameter $D_f$, flow rate $Q$, and the gravitational acceleration $g$. Based on the above parameters, the film dynamics is described by four non-dimensional  parameters, namely, Kaptiza, $\Gamma= \sigma/(\rho \nu^{4/3} g^{1/3}) = (l_c/l_{\nu})^2$ which is the ratio of the capillary forces and viscous forces. Here,  the capillary length scale $l_c= \sqrt{\sigma/(\rho g)}$ and the viscous length scale $l_{\nu} = \nu^{2/3} g^{-1/2}$ is obtained by a  balance between the gravitational and viscous forces. Using the inlet volumetric flow rate, $Q$, we can define the Reynolds number $Re = \rho Q/\mu \pi R_n$, $\alpha=R_n/R_f$ and Bond number $Bo=\rho g R_n^2/\sigma$. We note that  \cite{duprat2009spatial} Nusselt film thickness $h_N$ in contrast to the nozzle radius, $R_n$, used here.

In the present study, we investigate the role of $Re$ on the film dynamics keeping the other nondimensional numbers constant. Reynolds number, $Re$, is varied over from 0.13 to 2.4. For the experiments in the present study, based on the properties Silicon oil (see above), the other nondimensional number are given by Kaptiza number, $\Gamma = 0.83$, and Bond number, $Bo =0.27$. The nozzle to fibre diameter ratio is fixed at $D_n/D_f = 14.28$. In order to understand the effect of corrugations on the fiber on the size and velocity of the fluid beads, we perform experiments using fibers with a circular solid bead placed several fiber diameters downstream of the nozzle. We characterize such fibers for film flow dynamics by defining a parameter, blockage ratio ($BR$), which is similar to the flow blockage ratio used in channel flows with bluff bodies.\\
First, we perform experiments without the solid bead, to see the effect of the flow rate on the trains of waves generated on the film. Figure \ref{fig:flatfibre_events} shows a regime transition, that is change in the geometry, spacing and speed of the fluid beads, as a function of $Re$. Figure \ref{fig:flatfibre_events}a shows formation of beads with large spacing ($\lambda_b = 75mm$) with ripples in the intermediate fluid film for $Re = 0.13$. The nozzle is 25mm above the region depicted in the figure. The corresponding spatio-temporal graph \ref{fig:flatfibre_events}b, obtained by summing the intensity graphs in the transverse direction (see Appendix), shows that the fluid bead spacing is nearly uniform and the ripples undergo continuous merging with the travelling fluid beads. As the droplets pass they also leave behind rippled interface and overall they do not accumulate material and drops move at constant speed. The fluid bead speed is 138 mm/s indicated by the slope of the crest lines on the spatio-temporal graph. The bead shapes are similar to the droplike regime discussed in \cite{duprat2009spatial}. With increase in $Re$ to 0.155, we observe a decrease in $\lambda_b$ to 9 as shown in the figure \ref{fig:flatfibre_events}c. We also note that all the fluid beads even at higher $Re$ are separated by a uniform distance and travel at a constant speed of 64 mm/s as calculated from the spatiotemporal graph shown in \ref{fig:flatfibre_events}d. The shape of the beads is similar to that for $Re = 0.13$, but the ripples on the film surface are not observed (no separate streaks observed in the spatio-temporal graph) and there is no change in the size of the beads as they travel downwards.  With a further increase in $Re$ to 0.22, we observe a substantial change in the film dynamics (see \ref{fig:flatfibre_events}e). Irregular waves are observed but the shape of the droplets indicate that they are in the droplike regime. This irregularity is an indication of the transition regime between Rayleigh-Plateau and the Kapitza instability as discussed in \cite{duprat2007absolute}. The primary instability is convective but the dominant instability is characterized by Rayleigh Plateau instability (as also discussed in \cite{duprat2009spatial}. We observe a coarsening of the droplet as a fast moving droplet sweeps over several droplets resulting in a single big droplets as shown by the spatio-temporal graph in \ref{fig:flatfibre_events}f. This regime was named as droplet merging regime \cite{duprat2009spatial,ji2020modelling}. 
 
    \begin{figure}
  \centering
   \captionsetup{width=\linewidth}
  \captionsetup{justification=justified} 
  \includegraphics[width=0.8\textwidth]{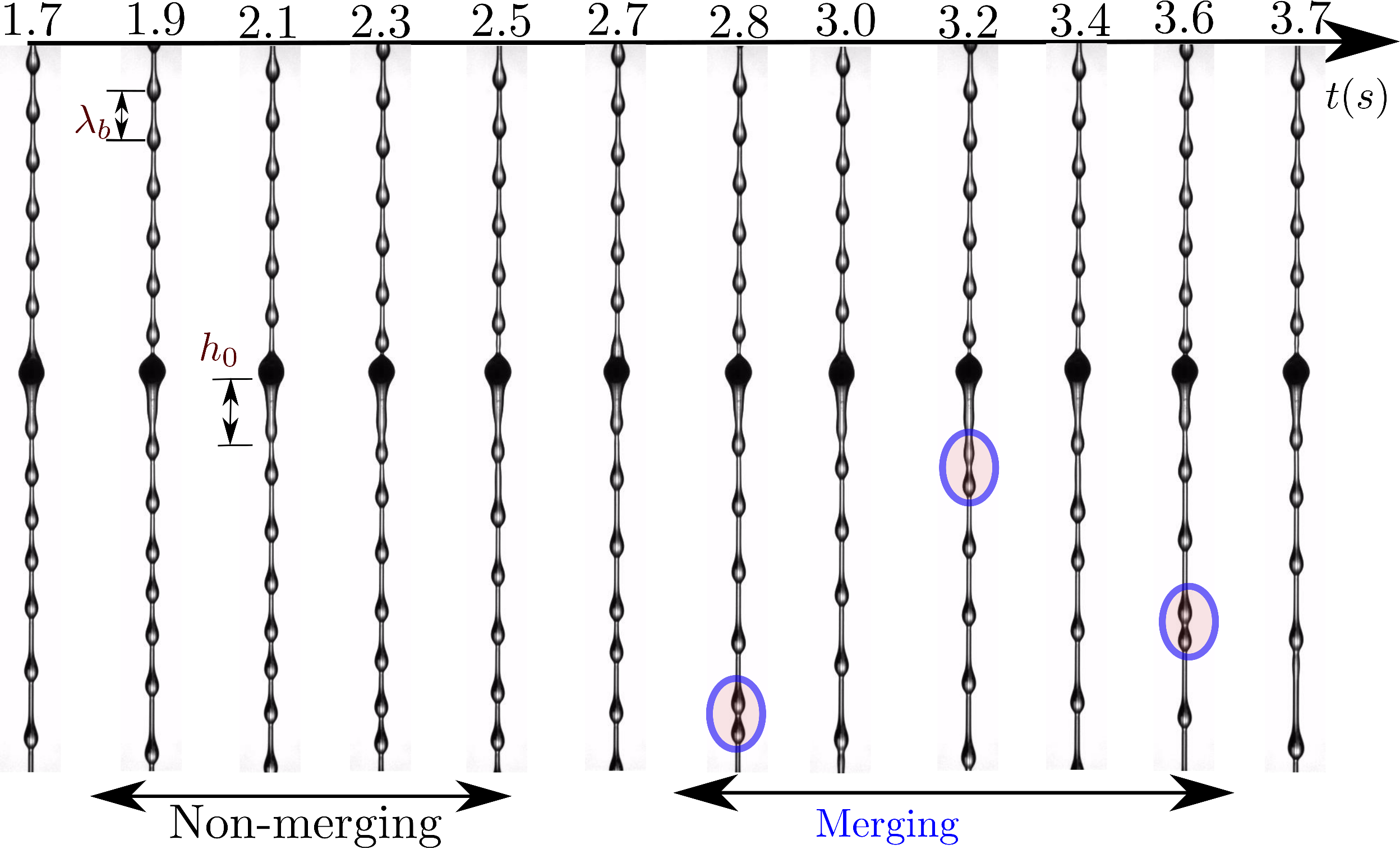} \caption{\label{fig:exp_bead_fiber} The instantaneous profiles of the film growth over the nozzle to fibre $D_n/D_f=14.28$ with bead ratio $BR=0.88$ at $Re=0.155$ are displayed with timestamps. A periodic solution (Rayleigh Plateau) regime over the obstruction is switched out by a droplet merging regime downstream of the obstruction. The transition occurred with a healing length generated below the bead, which appears to reduce the time delay in the droplet formation, causing the droplet to form at $t=2.8$ and repeated at $t=3.2$ and $t=3.6$ in the merging regime. The time gap between any two images is $0.18$s.}
\end{figure}

  \begin{figure}
  \centering
\captionsetup{width=1\linewidth} 
\captionsetup{justification=justified}\includegraphics[width=1\textwidth]{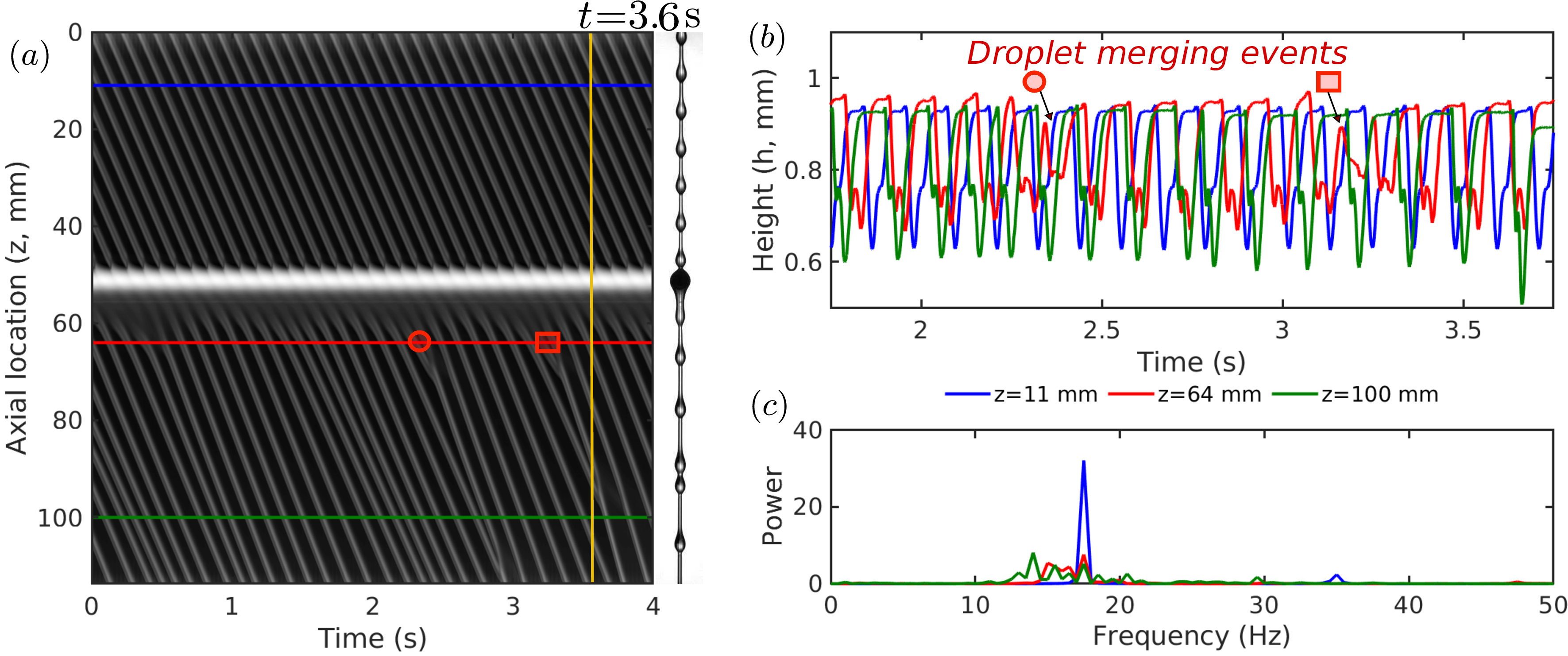} \caption{\label{fig:spatiotemp_exp_bead} (a) The spatio-temporal evolution of the film is shown  for $Re=0.155$ which is in the Rayleigh Plateau regime before the blockage and droplet merging down stream of the blockage. Red line on the spatio-temporal map is at time instance  at $t=3.6$ seconds whose interface evolution is shown next to the map. (b) The film thickness is shown at two different axial locations $11$ mm and $64$ mm showing the influence of blockage on the film thickness at given location and (c) shows the film thickness dominant frequency peak of 18.5 Hz. However, in the downstream direction, the peaks spread out into multiple peaks.}
\end{figure}

We obtain the above regimes by progressively increasing the flow rate over a fiber with uniform radius. If the flow is over a granular chain, we expect a shift in the critical flow rates at which these regime transition is observed. In order to study the effect of the beads of a granular chain on the formation of droplets, first we investigate the effect of a single rigid bead on the fiber $\sim 25mm$ downstream of the nozzle. Figure \ref{fig:exp_bead_fiber} shows the evolution of the fluid film on a fiber with a bead. Upstream of the bead, the droplets initial form in the Rayleigh Plateau regime with regular spacing. However, once the droplets so formed cross over the blockage (solid bead), irregular spacing and droplet speeds are observed and subsequently several droplet mergers occur in the downstream at irregular intervals. This transition from regular wave patterns to irregular waves is similar to that described by Duprat et al. \cite{duprat2007absolute} as absolute to convective instability transition for a uniform fiber on increasing the flow-rate. As marked in the figure \ref{fig:exp_bead_fiber} corresponding to the time-stamp ($t = 0.83$), the droplet approaching the solid bead slows down and forms an elongated structure, similar to the healing length at the nozzle. Subsequently, a droplet forms as shown in the time $t=0.76$s to $0.97$s. The droplet so formed detaches and merges with previously detached droplet. This sequence of events is similar to the irregular droplet formation from faucet in the chaotic regime. We note that droplet merger events also occur irregularly and between merger events several isolated droplets form. The spacing between droplets sufficiently downstream of the bead is $\sim 10$mm in-contrast to spacing between the droplets upstream of the blockage $\sim 5.8$mm .  
\begin{figure}
  \centering
\captionsetup{width=1\linewidth} 
\captionsetup{justification=justified}\includegraphics[width=1\textwidth]{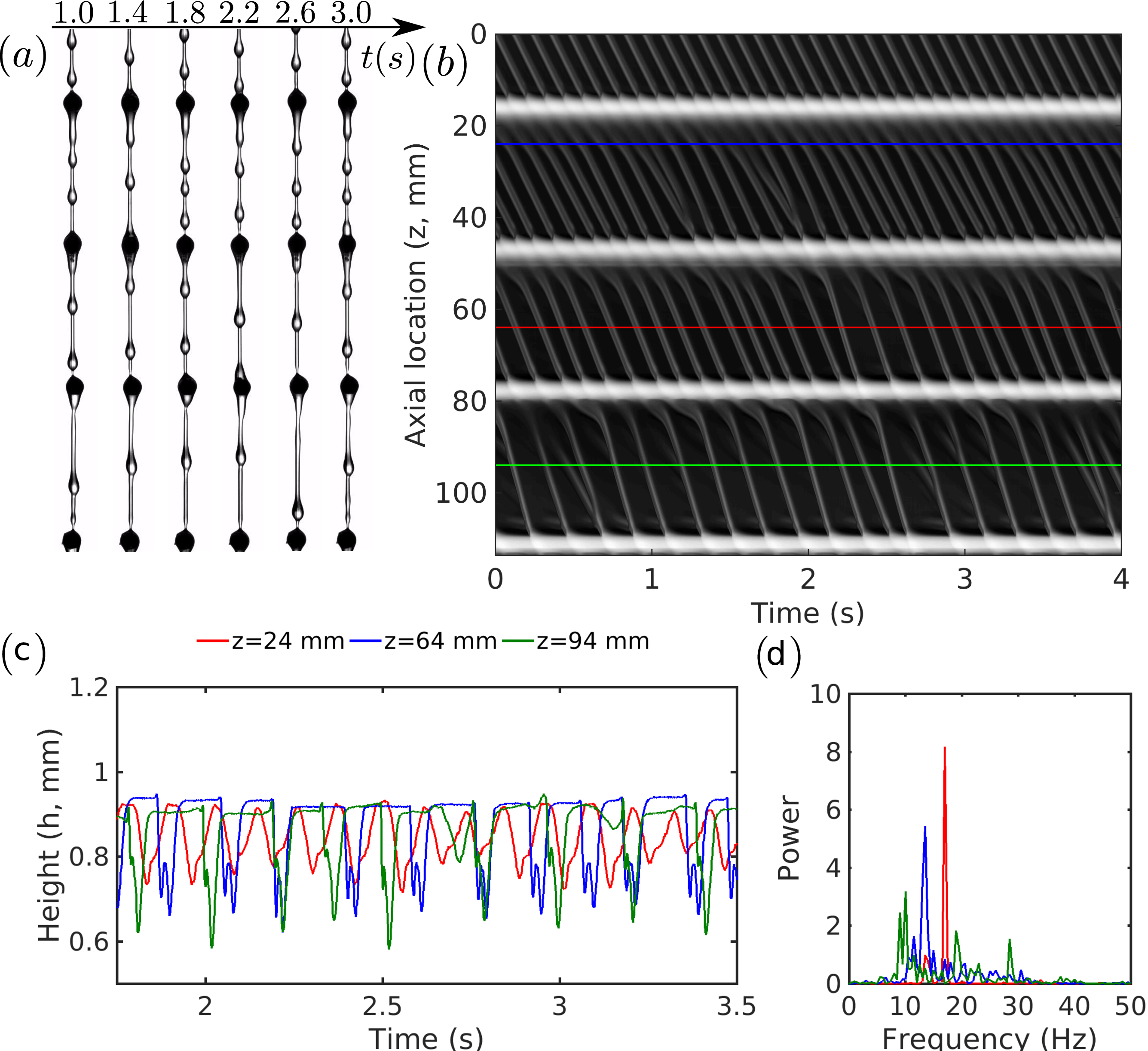} \caption{\label{fig:multibead_25mm_spacing} (a) Droplet motion along the fibre with bead spacing 25 mm at different instant of time.The interval between any two images in the figure a is $0.4$s (b) The spatio-temporal evolution of the film is shown  for $Re=0.155$ which is in the Rayleigh Plateau regime before the blockage and droplet merging down stream of the blockage.}
\end{figure}
\begin{figure}
  \centering
\captionsetup{width=1\linewidth} 
\captionsetup{justification=justified}\includegraphics[width=1\textwidth]{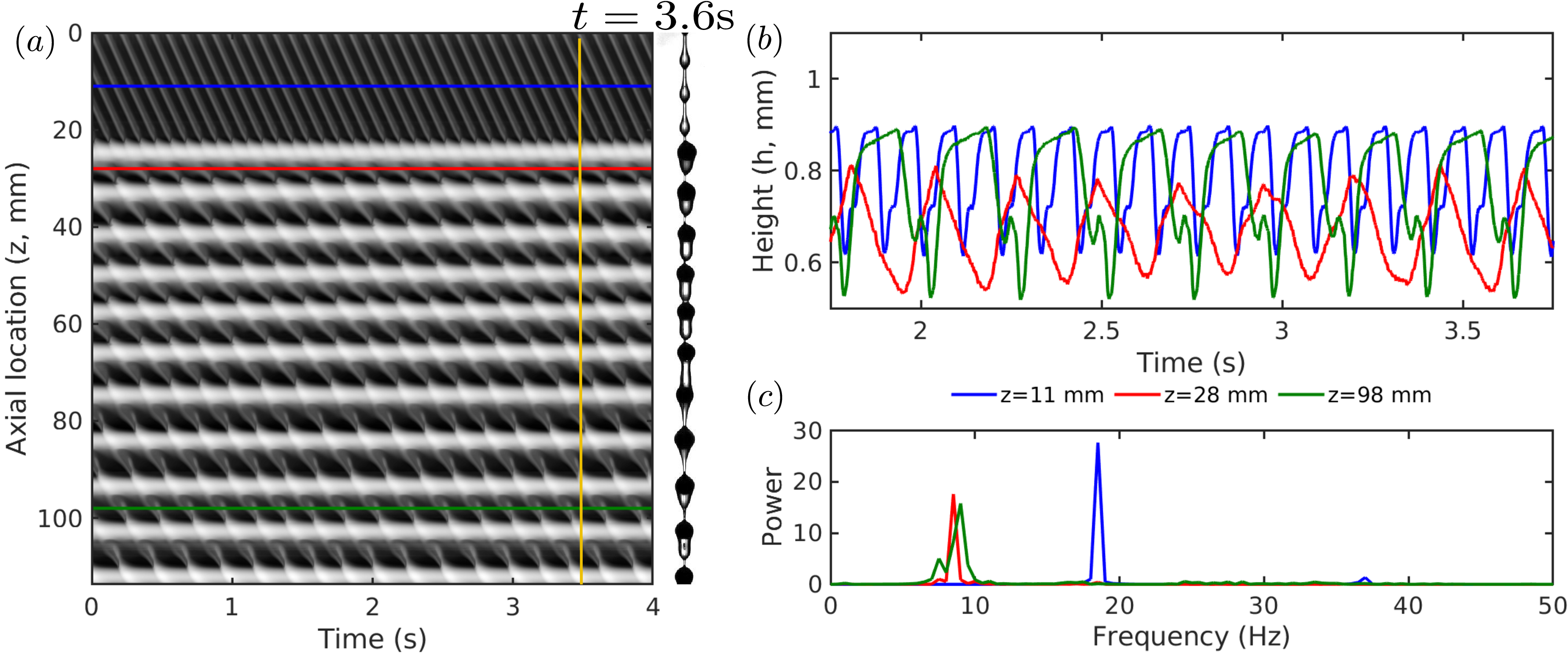} \caption{\label{fig:multibead_7mm_spacing_exp} (a) Droplet motion along the fibre  with bead spacing 7 mm  at different instant of time (b) The spatio-temporal evolution of the film is shown  for $Re=0.155$ which is in the Rayleigh Plateau regime before the blockage and droplet merging down stream of the blockage.}
\end{figure}

Figure \ref{fig:spatiotemp_exp_bead} shows the spatio-temporal map corresponding to the time evolution of the interface profiles shown in figure \ref{fig:exp_bead_fiber}. The solid yellow line marked on the spatio-temporal plot corresponds to the interface profile at $t = 3.6$s shown adjacent to figure \ref{fig:spatiotemp_exp_bead}a. The solid bead is located at $45$ mm axial location on the fibre, and corresponds to the bright static area on the spatio-temporal map. As the fluid flows over the blockage, the film flow slows down resulting in the formation of an elongated thicker region instead of a droplet just upstream of the bead as shown in figure \ref{fig:exp_bead_fiber}a. We note here that although the incoming droplets are uniformly spaced and move with constant speeds (suggested by the slope of the crest lines in the spatio-temporal map), the droplets in the downstream of the blockage form at irregular intervals with irregular spacing and move with different speeds as indicated by the variation of film thickness ($h$) with time as shown in figure \ref{fig:spatiotemp_exp_bead}b. The FFT of the time series in figure \ref{fig:spatiotemp_exp_bead}c shows that the introduction of the blockage broadens the spectrum peak. The power spectrum for the upstream location at $z = 11mm$ has a single dominant peak in comparison to that for the downstream location $z = 64mm$.  This transition is similar to that observed in the drop merging regime shown in figure \ref{fig:flatfibre_events}e and f observed for higher $Re = 0.22$ (also see figure 6 in \cite{duprat2009spatial}).
When multiple blockages are placed at a uniform spacing of $25$mm (where the spacing between the droplet upstream of the beads is $\lambda_b \sim 9$), as shown in figure \ref{fig:multibead_25mm_spacing}a, the droplets are seen to further accelerate from $62.3$ mm/s to $70$ mm/s (computed using the slopes of the crest line in Fig.\ref{fig:multibead_25mm_spacing}b) as they travel after the second blockage to third blockage and continue to accelerate after every blockage along with an decrease in the spacing between the droplets before droplet merging event. After the droplet merger, there increase in the spacing between the droplets, similar to the case for low $Re \sim 0.155$, we observe intermittent formation of ripples corresponding to Rayleigh-Plateau instability. Figure \ref{fig:multibead_25mm_spacing}c, shows variation in height of the film with time at three different locations marked in the spatio-temporal plot in Fig.\ref{fig:multibead_25mm_spacing}b. The corresponding FFT is shown in Fig.\ref{fig:multibead_25mm_spacing}d. The FFT shows broadening of the peak after the first blockage similar to the one for the single bead in figure \ref{fig:spatiotemp_exp_bead}c (red curve). The FFT has signature of increase in irregularity (broadening of the spectrum) in droplet spacing in the downstream after the second and the third beads with the peak of the spectrum shifting towards left indicating a progressively decreasing velocity of the droplets. Interestingly, two superharmonic peaks appear in the spectrum corresponding to $z = 94mm$ which may be due to ripple formation as also seen in the spatio-temporal plot. 

When we substantially reduce the spacing between the beads, such that the droplet spacing in the upstream of the beads ($\lambda_b= 9$mm) is larger than the bead spacing ($7$mm), most interfacial features seen earlier are not observed. Since, the critical wavelength for the Rayleigh-Plateau instability is $\sim 9.4mm$, we do not observe droplet formation, instead the fluid drains through the formation of fluid filaments attached to the beads as seen in figure \ref{fig:multibead_7mm_spacing_exp}a. Interface profile at $t=3.55$s, indicated with a yellow vertical line in figure \ref{fig:multibead_7mm_spacing_exp}(a), is plotted adjacent to it. The slope of the crest lines in the plot, clearly shows a substantial reduction in the speed of the droplet motion from   \( 62.3 \, \text{mm/s} \) to  \( 18.3 \, \text{mm/s} \). Interesting, unlike the case where spacing between the beads is larger, for the $7mm$ bead spacing case the speed of the traversing droplets remains the same as they pass over a cascade of solid beads. Evolution of the film thickness just upstream and downstream of the first bead (at $z=11$ and $z=28$mm) and downstream of the ninth bead (at $z = 98$mm) are shown in figure \ref{fig:multibead_7mm_spacing_exp}b with the corresponding FFT shown in figure \ref{fig:multibead_7mm_spacing_exp}c. The FFT shows that the initial motion of the droplet formation is significantly reduced to $\sim$half of the upstream droplet formation frequency. Moreover, the FFT for $z=28$mm and $z=98$mm are nearly superposed. 

 In order to further investigate the effect of solid beads (size and location) on film flow and droplet formation, we perform detailed numerical simulations. We note that, the numerical formulation is first presented and the experimental condition presented and \cite{ji2020modelling} are compared to further understand the effect of torus geometry affect on the film growth. This study ends with granular chain whose influence on the film growth is presented. 
\begin{figure}
\centering
\captionsetup{width=1\linewidth}
\captionsetup{justification=justified}  \includegraphics[width=0.6\textwidth]{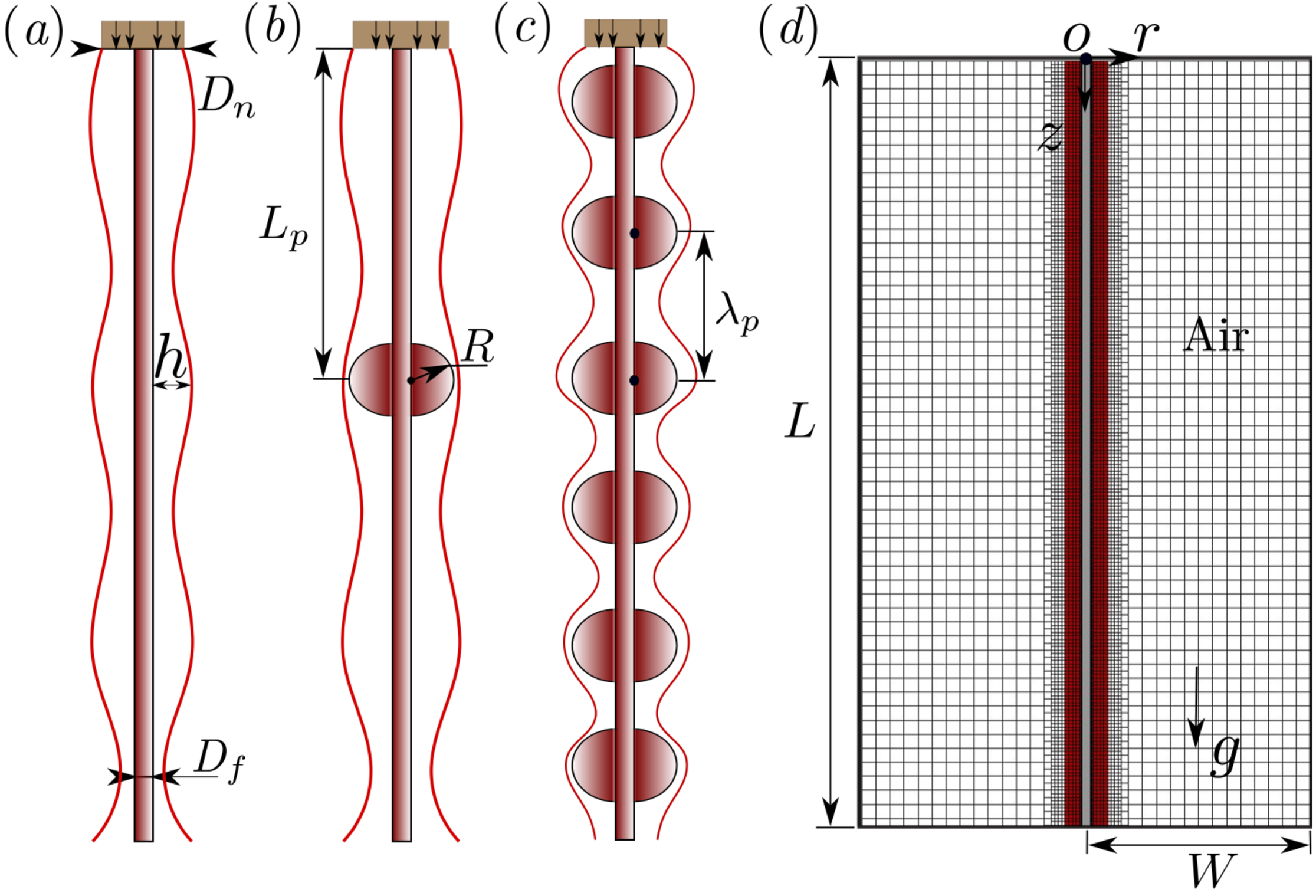}
\caption{ $(a)$ A  local fluid film  with thickness denoted by $h$ flows over a cylindrical fibre of diameter $D_f$ from a nozzle of diameter $D_n$. $(b)$ A fibre with rigid torus bead of radius, $R$ located at $L_p$ from the nozzle is exit  $(c)$  a fibre with torus granular chain with spacing between them $\lambda_p$ is shown and $(d)$ Computational domain of length $L$ and width $W$. A liquid film of thickness $h$ enters the computational domain from above. A coordinate frame ($r-z$ plane) is attached at the axis of symmetry with $z$-coordinate pointing in the direction of the gravitational acceleration ($\mathbf{g}$). Ambient fluid is air (with density $\rho_g = 1kg/m^3$) and is assumed to be quiescent. Outflow boundary conditions are imposed at the bottom end.
  }
\label{fig:Domain}
\end{figure}

\begin{figure} 
  \centering
   \captionsetup{width=\linewidth}
  \captionsetup{justification=justified} 
  \includegraphics[width=1\textwidth]{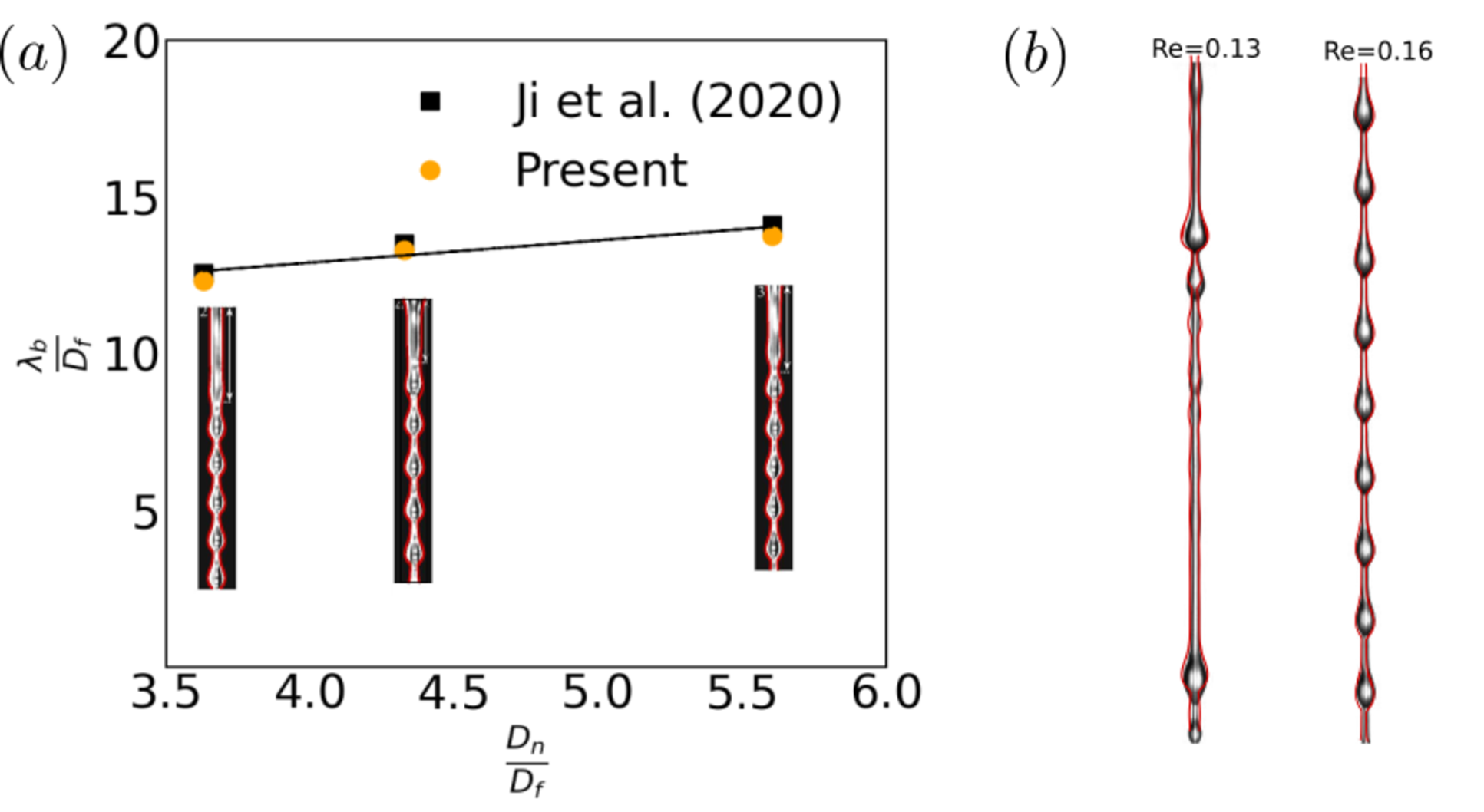} 
 \caption{\label{fig:Validation} (a) Comparison of numerical simulations (filled black circles)  with the experimental results  of \citet{ji2020modelling} (filled orange squares).  
 Variation in the bead spacing ($\lambda_b/D_f$) with nozzle diameter ($D_n/D_f$) is shown for three different configurations. Corresponding interface morphology have also been compared. Interface profiles obtained from numerical simulations (marked in red) are overlaid on the experimental images. Note that the dashed lines between the points are only suggestive and do not indicate the exact behavior. (b) Present experimental conditions for $D_n/D_f = 14.28$ compared with the DNS results with increase in the Reynolds number. } 
\label{gridindependence}
\end{figure} 

 \begin{table}
  \begin{center}
\def~{\hphantom{0}}
  \begin{tabular}{lccc}
        $D_n/D_f$  & $Re$ & $Bo$   & $We$  \\[3pt]
                     
        3.63 & 1.2 & 1.1 & 0.12 \\
        4.33 & 0.9 & 1.5 & 0.059 \\
        5.60 & 0.8 & 2.6 & 0.032
  \end{tabular}
  \caption{Non-dimensional parameters  calculated based on the estimated velocities for the validation of experimental results \citep{ji2020modelling}}
  \label{tab:ji_experimental}
  \end{center}
\end{table}

\section{ Numerical formulation and observations } \label{Numerical}
  
 We numerically study the dynamics of a Newtonian liquid film (density $\rho_l$ and viscosity $\mu_l$) in quiescent air (density $\rho_a$ and viscosity $\mu_a$) flowing down from a nozzle axisymmetrically over a thin fiber. Schematics given in figures \ref{fig:Domain}a,b, and c show the various configurations we investigate in this study. Figure \ref{fig:Domain}a shows the configuration for a flat fiber which has been studied earlier in \cite{ji2020modelling} for different nozzle to fiber diameters, $D_n/D_f$, where $D_n$ and $D_f$ are the nozzle and fiber diameters, respectively. The local thickness of the fluid film is denoted by $h(z,t)$ which is a function of the axial coordinate $z$ and time $t$. Radial coordinate is denoted by $r$ and the origin is placed at the center of the nozzle. Figure \ref{fig:Domain}b shows the configuration we employ to study the effect of the presence of a corrugation on the fiber surface on the flow instability characteristics. The corrugation is modeled using an axi-symmetric particle with radius of curvature $R$ and is placed at a distance $L_p$ from the nozzle. Figure \ref{fig:Domain}c shows the geometry employed to study flow over a granular chain. The granular chain is made up of solid beads with a radius $R$ placed at a spacing of $\lambda_p$. \\
 
 We perform numerical simulations using the volume-of-fluid method in the two-phase flow open source code Gerris (\cite{popinet2003gerris,popinet2009accurate,tomar2010multiscale}). Computational domain chosen for the simulations is shown in figure \ref{fig:Domain}d. We perform axisymmetric simulations and thus only half of the domain, defined by length $L$ and width $W$,  shown in the figure is employed for performing the simulations. The gravitational acceleration is acting in the $z$ direction (downward) and the ambient fluid is air. The computational domain is initialized with a flat film of liquid of uniform thickness ($D_n - D_f$). Different geometries shown in the schematics in figures \ref{fig:Domain}a,b and c are generated using the following functions. For the flat cylindrical fiber, 
 \begin{equation} \label{cylindrical}
 f(z) = R_f.
\end{equation}
where $R_f$ is the radius of the fiber. For the configuration in  figure\ref{fig:Domain}b, the following functional form is used to define the fiber and the corrugation:
\begin{equation} \label{single_torus}
    f(x)= 
\begin{cases}
    R_f+[R^2-(z-L_p)^2]^{1/2} ,& \text{if } z-R \leq z\leq z+R \\
    R_f,              & \text{otherwise.}
\end{cases}
\end{equation}
A granular chain geometry (Figure \ref{fig:Domain}c) can be similarly created by introducing a spatial periodicity in the above function.
Since, the solid bead obstructs the flow, we define a geometric parameter, bead ratio, to indicate the degree of bead to the liquid flow. bead ratio can be defined as:
\begin{equation} \label{eq.torus}
      BR=\frac{2R}{D_n-D_f}.
\end{equation}
where $R$ is the radius of curvature of the solid bead on the fiber.

 The Gerris flow solver employed in the present study is based on a one-fluid formulation. Both the liquid and the ambient air are considered incompressible. The Navier-Stokes equations for momentum conservation are scaled with the nozzle diameter chosen as the characteristic length ($D_n$) and the inlet velocity ($v$) chosen as the velocity scale. The Navier-Stokes equation is modified to include the surface tension force term (with surface tension coefficient $\sigma$) acting on the interface (embedded in the Eulerian grid) and is expressed as a volumetric force using a surface Dirac delta function ($\mathbf{\delta_{s}}$) using the continuum surface force (CSF) model \cite{brackbill1992continuum}. The resulting non-dimensionalized equations for incompressibility and momentum conservation are respectively given as, 
    \begin{equation}
      \nabla \cdot \mathbf{u}=0 ~~\mbox{and}
  \label{continuity}
  \end{equation}

    \begin{equation}
      \rho \left[ \frac {\partial \mathbf{u}}{\partial t}+\left( \mathbf{u}\cdot \nabla \right)  \mathbf{u}\right] =-\nabla P+\left( \frac {\kappa}{We}\right)\mathbf{n}\delta_{s} -\rho \left( \frac {Bo}{We}\right) \\+\left( \frac {1}{Re}\right) \nabla \cdot \left[ \mu \left( \nabla \mathbf{u}+\nabla \mathbf{u}^{T}\right) \right]
       \label{NS}
\end{equation}
where $\kappa$ and $\mathbf{n}$ are the curvature and the unit normal at the interface, respectively. The non-dimensional numbers governing the problem are, Weber number $We = {\rho_l v^2 D_n}/{\sigma}$, Reynolds number $Re = {\rho_l v D_n}/{\mu_l}$ and Bond number $Bo = {\rho_l g D_n^2}/{\sigma}$. The density ($\rho$) and viscosity ($\mu$) in the above equation (Eq.\ref{NS})are scaled by the liquid properties and are a function of the local liquid fraction (ratio of the liquid volume in a given Eulerian grid cell to the volume of the grid cell), $\alpha$,    
\begin{equation} 
    \rho =\alpha +\left( 1-\alpha \right) (\rho _{a}/\rho _{l})
\end{equation}
\begin{equation}
     \mu =\alpha +\left( 1-\alpha \right) (\mu _{a}/\mu _{l})
\end{equation}

 The density ratios $\rho_a/\rho_l=1.5 \times 10^{-3}$ and $\mu_a/\mu_l=3.7 \times 10^{-4}$ are fixed based on the liquid properties given in the experimental study by \cite{ji2020modelling}. The interface evolution is captured by solving the advection equation for volume fraction $\alpha$ using geometric volume of fluid method,
\begin{equation}
         \frac {\partial \alpha }{\partial t}+ \boldsymbol{u}\cdot\nabla\alpha =0
\end{equation}

 A uniform flow with velocity $v$ from the nozzle region ($D_n$) on the top surface is imposed as the inflow condition. A  no-slip boundary condition is imposed on the fiber surface. At the bottom surface of the domain an outflow boundary condition is imposed, that is, a Dirichlet condition on the pressure ($P=0$), and a Neumann condition on the axial and radial velocities, given by $\partial_zU = 0$ and $\partial_zV =0$, respectively. Axi-symmetric boundary conditions are imposed on the $r=0$ surface and slip boundary condition is imposed on the $r = W$ surface. The length of the computational domain corresponds to $L = 87mm$ and $W = 10mm$. The adaptive mesh refinement is employed with the interface being resolve using a fine mesh size of $\delta=4.7 \mu m$, whereas a coarser mesh ($\delta=37 \mu m$) is employed in the gas region away from the interface. These mesh sizes yield grid independent results.\\  
 \noindent We first validate our numerical formulation using the experiments in  \cite{ji2020modelling} for the  configuration in figure \ref{fig:Domain}a corresponding to a film flow over a flat fiber. Numerical simulations have been performed for three values of $D_n/D_f = 3.63$, $4.33$ and $5.60$ that correspond to different $Re$, $Bo$ and $We$ numbers given in table \ref{tab:ji_experimental}. Figure \ref{fig:Validation} shows the variation in the spacing between the fluid beads on the fiber, $\lambda_b/D_f$, with increasing $D_n/D_f$ that are in good agreement with the experimental observations of \cite{ji2020modelling}. The corresponding interface profiles are presented in the insets given in Fig.\ref{fig:Validation}. The interface profile from the numerical simulations are shown in red and are drawn over the experimental observations from \cite{ji2020modelling}. This clearly shows that several features of the instability of fluid flow on a fiber, such as healing length (length of the fiber over which instability develops to form beads; see \cite{duprat2007absolute}), amplitude and wavelength of the fluid beads, and their variation with increase in $D_n/D_f$, are captured accurately using the current numerical formulation. In the reduced order numerical model proposed in \cite{ji2020modelling}, a van der Waals based repulsion term (for wetting fluids), using a stabilizing Hamaker constant, was employed to prevent the film thickness to become negative (that is to prevent the film profile to penetrate the fiber). For this a critical minimum film thickness ($\epsilon_p$) was chosen and a corresponding critical value of the Hamaker constant was derived using the stability condition for a stable coating of thickness $\epsilon_p$. A sensitivity analysis presented in \cite{ji2020modelling} showed dependence of bead spacing and bead velocity on the choice of $\epsilon_p$ and an appropriate value of $\epsilon_p$ was required for different flow rates. Typical values of $\epsilon_p$ chosen was in the range 0.15-0.3mm, which correspond to a thickness much larger than the thickness of $100nm$ above which van der Waals forces are known to have little effect on the dynamics of the film \cite{oron1997long}. In contrast to that, in the Navier-Stokes based model used in the present study, we did not employ any such model parameter. The thin regions are sufficiently resolved using $256$ number of grid points in the thinnest regions.  
\begin{figure} 
  \centering \captionsetup{width=\linewidth} 
  \captionsetup{justification=justified} 
 \includegraphics[width=1.0\textwidth]{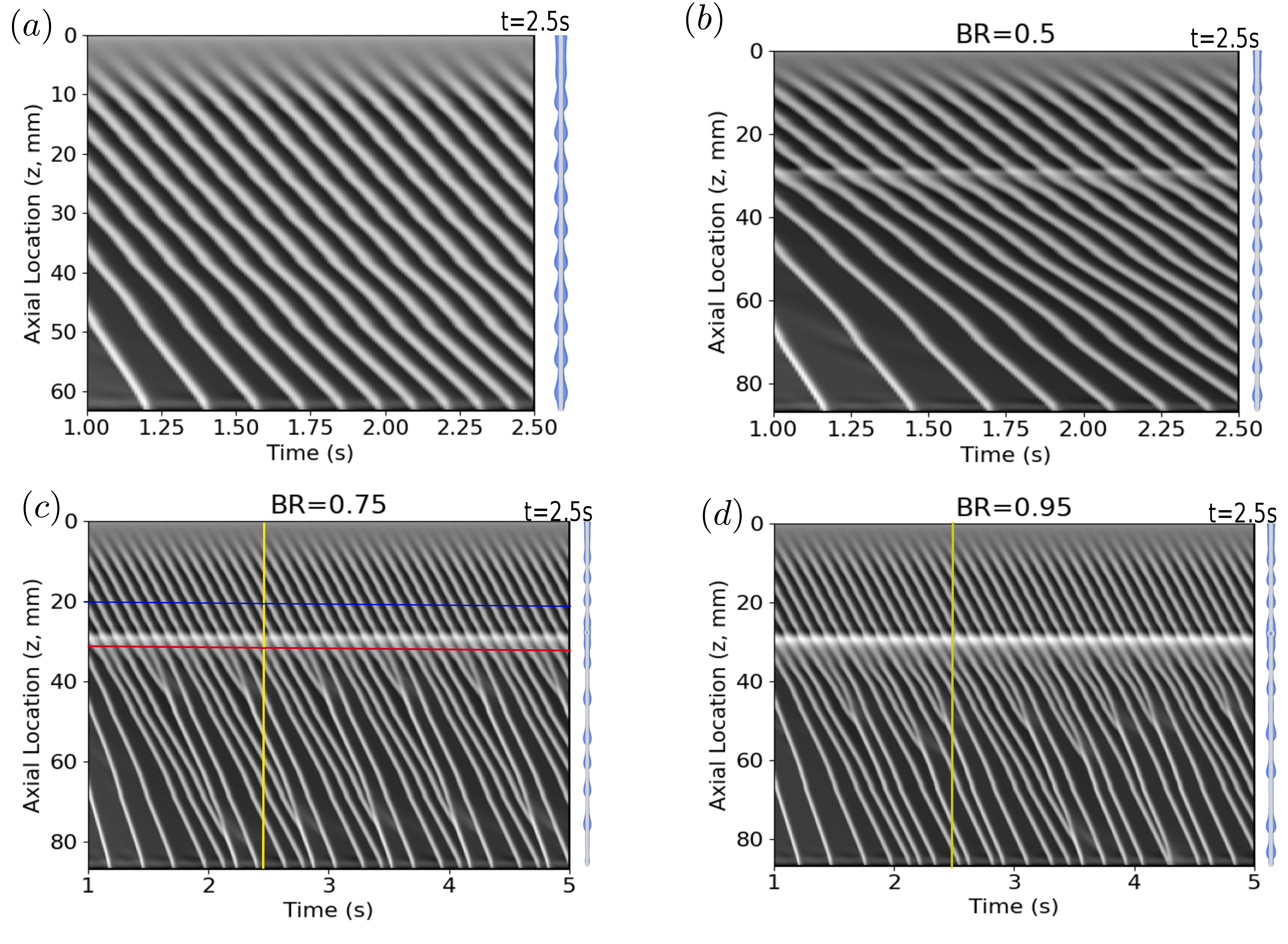}
    \caption{ \label{fig:BR_influence} space-time graph indicating the motion of the liquid beads on fibre for different bead ratios (a) $BR = 0.0$ (b) $Re = 0.5$  (c) $BR = 0.75$ and (d) $BR = 0.95$. Other parameters used for the numerical simulations are provided in Table \ref{tab:Blockageratios}. Corresponding interface profiles at $T_0 = 2.50 \text{s}$ have also been plotted. dark contours imply thinner regions whereas white contours mark regions of maximum thickness of the fluid film, essentially marking the axial locations of the liquid beads on the fibre.}
\end{figure}   
\subsection{Results and discussion} 
In this section, we present Blockage Ratio (BR) and multiple blockages affect on the liquid film. The parameters used for the simulation are listed in the table \ref{tab:Blockageratios}. The shape of the blockage is defined as a torus which is specified by the equation \ref{single_torus}. This geometry closely mimics the experimental geometry is defined and visually illustrated in the figure \ref{fig:Domain}(b). The Rayleigh-Plateau instability, observed on the fibre with $BR=0$ at $Re=0.9$ shown in the figure \ref{fig:BR_influence}(a) which is experimentally studied by \cite{ji2020modelling} is considered as basis for the simulations. The figure shows space-time graph with healing length ($h_0$) near the nozzle and uniformly spaced wave crest (droplets) in the entire time domain of the simulation with $\lambda_b=5.6$. The interface profile is shown right to the figure \ref{fig:BR_influence}(a) at $t=2.5$s. At this specific $Re=0.90$, we focus on the role of fibre geometry (torus blockage) size and position on the film evolution. In the figure \ref{fig:BR_influence}(b-d), the position of the bead is located at $30D$ downstream from the nozzle exit, allowing the droplet formation similar to experimental conditions. 
   
\begin{table} 
\centering
\begin{tabular}{cllllll}
\textbf{Bead Ratio ($BR$)} & 0 & 0.5 & 0.75 & 0.95
\end{tabular}
 \caption{\label{tab:Blockageratios} Effect of bead Ratio ($BR$) for torus geometry are studied using different $BR$ values given here at $D_n/D_f=4.32$ in the Rayleigh Plateau regime $Re=0.90$. The initial observations at $Re=0.90$ are discussed in the section \ref{Numerical}. }
\end{table}
\begin{figure} 
  \centering
   \captionsetup{width=\linewidth}
  \captionsetup{justification=justified} 
  
  \includegraphics[width=1\textwidth]{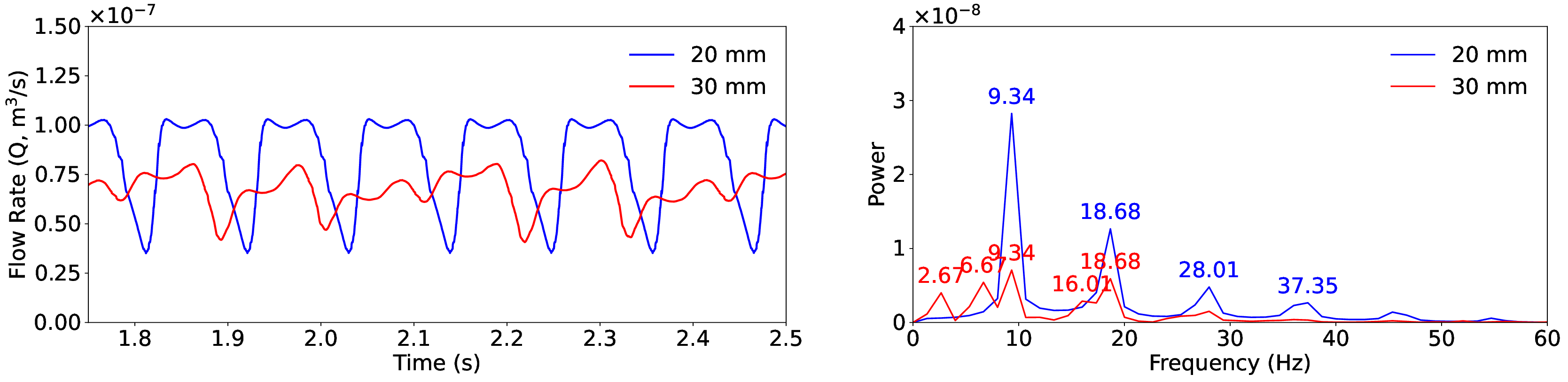}
  \caption{\label{fig:flowrate_variation} (a) The temporal evolution of flow rate (Q) are shown at two axial locations-upstream blockage at $z=20$mm and near the blockage at $z=30$ mm. (b) The influence of the blockage is shown with  frequency distribution spread highlighting the frequency values with harmonic and harmonic peaks.}
  \end{figure}  
Figure \ref{fig:BR_influence}(b) shows evolution of the fluid film on a fibre with $BR=0.5$. Upstream of the bead, the droplets initially form in the Rayleigh Plateau regime with regular spacing, similar to the observations seen in the experimental conditions. However, once the droplets so formed cross over the bead, initially the bead has a coarsening effect and subsequently droplets have attained uniform spacing. Unlike the experimental observations, downstream of the bead has not developed healing length observed in experimental conditions. The film evolution on the fibre captured at $t=2.5$s doesnot show the evolution of elongated structure. Figure \ref{fig:BR_influence}(c) shows the spatio-temporal map corresponding to $BR=0.75$. The upstream has seen any influenced on the film evolution similar to $BR=0.5$ and experimental observations. However, the bead influence on the film in the downstream is clearly evident. The film has started showing healing length ($h_0$) and droplet merging corroborating the experimental observations shown in figure \ref{fig:spatiotemp_exp_bead}. The simulations are extended to longer time i.e. $t=5.0$s to understand the downstream droplet behavior. These spatio-temporal map indicates, periodic merging of droplet after every third droplet at a fixed location. Further increase in the $BR$ to $0.95$ is shown in the figure \ref{fig:BR_influence}(d). The downstream film evolution shows oscillatory behavior in the elongated structure, consequently the droplet merging has become chaotic and the crest lines shows reduced speed of the droplets(slope of the crest lines). The droplet that is crossing the bead is elongated on the bead is shown right to the space-temporal map at $t=2.5$s. 
 
\begin{figure} 
  \centering
   \captionsetup{width=\linewidth}
  \captionsetup{justification=justified} 
  \includegraphics[width=1\textwidth]{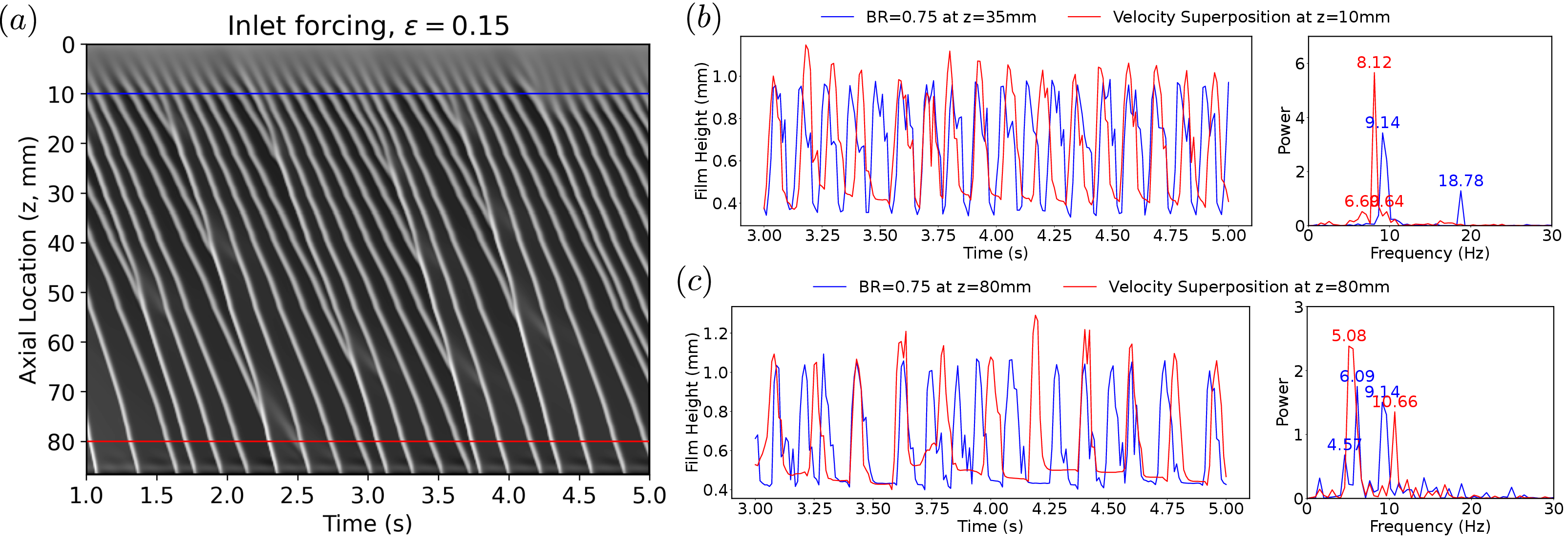}
 \caption{\label{fig: inlet perturbation} (a) Space time evolution of the film for inlet velocity perturbation by imposing the frequency distribution observed at downstream location at $z=30$mm for $BR=0.75$. (b) and (c) showing film height evolution with time at the inlet condition $z=10$, $z=80$. }
  \end{figure}  
  
In Figure \ref{fig:flowrate_variation}(a), we compare the flow rates at two axial locations $Q_{20}$ and $Q_{30}$  calculated from the radially-averaged axial velocity \( Q_{\text{axial}} = 2 \pi \int_{0}^{h} u_{axial}(r, z) r \, dr \) at a position in the upstream and near the bead for $Br=0.75$. Here, \(h(t)\) would be the film height which varies with the time at the given axial location, \(u(r,z)\) is the axial velocity at radial position \(r\) and axial location \(z\). This figure indicates a significant reduction in the amplitude from $z=20$mm to $z=30$mm axial location which indicates a strong influence by the bead on the flow rate variation. At $x=20$, the FFT figure \ref{fig:flowrate_variation}b reveals dominating peaks at frequencies \( f_{20} = \{9.34, 18.68, 28.01, 37.35\} \, \text{Hz} \) and at $x=30$ shows   \( f_{30} = \{2.67, 6.67, 9.34, 16.01, 18.68\} \, \text{Hz} \) for \( Q_{30} \). We can also see that there is no phase shift in the flow rate variation. These variations in the frequencies with in the time window shows variation of the amplitude between the upstream and downstream positions. Additionally, subharmonic peaks are detected at frequencies \( f_{20\text{sub}} = \{4.67, 9.34, 14.01, 18.68\} \, \text{Hz} \) for \( Q_{20} \) and \( f_{30\text{sub}} = \{1.33, 3.34, 4.67, 8.00, 9.34\} \, \text{Hz} \) for \( Q_{30} \), suggesting periodic components at fractions of the dominant frequencies. This reduced amplitude variation in the flow rate may be influencing a transition of film behavior from the Rayleigh-Plateau regime to a droplet-merging regime. This variation may result in periodicity in the droplet merging in the down stream of the bead $BR=0.75$ indicated in the figure \ref{fig:BR_influence}(c).    

The regime transition from equally spaced droplets to droplet merging regime may be represented as a perturbation induced by blockage at a given spatial location to a periodically evolving liquid film. The induced perturbation grows in the same way as velocity perturbation induced at the inlet of the nozzle, observed phenomena can be explained as temporal growing instability. This can be represented as superposition of the frequencies that are appearing at the downstream location ($z=30$mm).  
\[
V = v_0 \left( 1 + \epsilon_0 \left( \sin(\omega_1 t) + \frac{A_2}{A_1} \sin(\omega_2 t) + \frac{A_3}{A_1} \sin(\omega_3 t) + \ldots \right) \right)
\]
Each term is scaled by a factor that relates its amplitude \(A_n\) to the amplitude \(A_1\) of the first frequency component. When the inlet is perturbed with the frequencies obtained at $x=30$ shown in the figure \ref{fig:flowrate_variation} in the above form, the system shows the behaviour droplet merging regime similar to the $BR=0.75$, as shown in the figure \ref{fig: inlet perturbation}(a). This figure shows the spatio-temporal evolution of the liquid film perturbed at the inlet of the nozzle with the superposition velocity. The film evolution shows the healing length evolution near the inlet of the nozzle and then the droplet merging is observed the similar to the perturbation induced by the bead. As the film flows down the fibre, it shows further merging of droplets with reduced slope in the crest lines. This indicate increase in the droplet speed, this encourages further droplet merging. The spacing between the droplet also coarsened, similar observations are discussed by \cite{duprat2009spatial}. Figure  \ref{fig: inlet perturbation}(b-c) shows comparison of the film evolution on the fibre with inlet perturbation forcing and the $BR=0.75$ at two axial locations at the inlet $z=10$mm and down stream of bead $z=35$mm and at the downsteam location i.e. $z=80$mm. At $z=10$mm the film shows dominating frequency at $9.34$Hz which coincides with the bead conditions at $BR=0.75$ of the figure \ref{fig:BR_influence}c. 

 \begin{figure} 
  \centering
   \captionsetup{width=\linewidth}
  \captionsetup{justification=justified} 
  \includegraphics[width=1\textwidth] {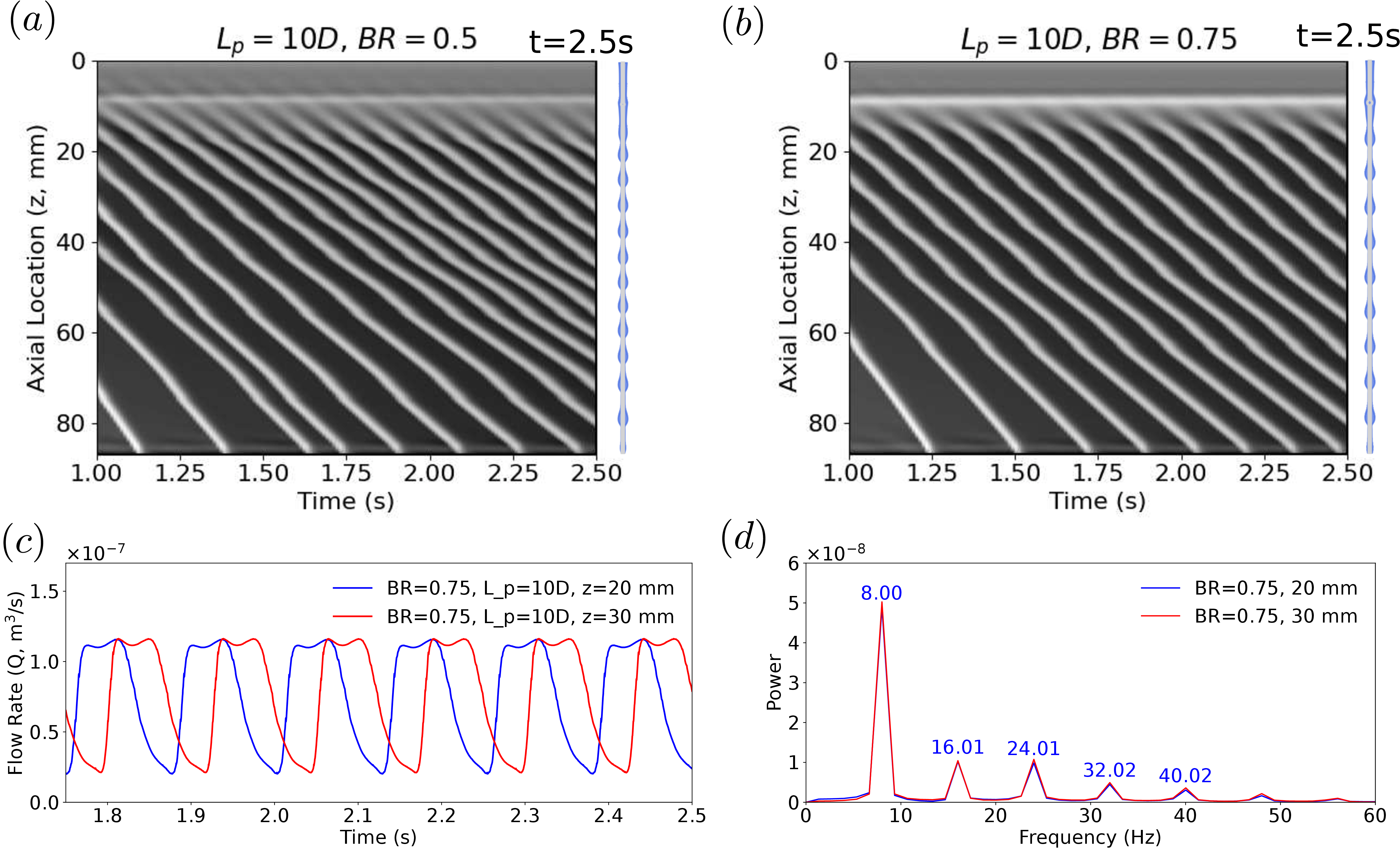} 
  \caption{  Effect of blockage location on the film evolution is shown for two BR ratios (a) BR=0.5 and BR=0.75. When both blockages are within the healing length of the liquid film near the nozzle, then the blockage has coarsening effect on the droplet spacing unlike the film evolution shown in the figure   }
\label{fig:BR_location}
\end{figure}
\begin{figure} 
  \centering
   \captionsetup{width=\linewidth}
  \captionsetup{justification=justified} 
  \includegraphics[width=1\textwidth]{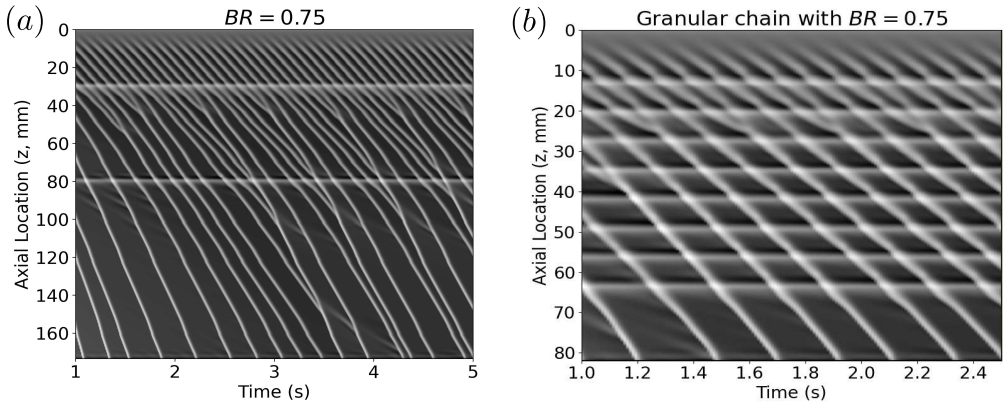}
  \caption{  Space -time graph of liquid beads on fibre with (a) two-solid bead, here computational domain chosen is $2L$ size (b) Granular chain of $BR=0.75$ at $Re=0.90$. Corresponding interface profiles shown at $T_0= 2.5 \text{s}$.     }
\label{fig:granular_chain}
\end{figure}

  
It is seen that beyond a critical $BR$, the bead has significant influence on the film evolution when it is located far from the nozzle exit ($L_p=30D$). Figure \ref{fig:BR_location}a,b,c,d shows film evolution when the bead location is within the healing length ($L_p=10D$) for $BR=0.5$ and $BR=0.75$. This configuration differs from the earlier configuration with the positioning of the bead. Under this condition, figure \ref{fig:BR_location}a shows space-time evolution with initial transitions for $BR=0.5$ at $L_p=10D$ similar to the $L_p=30D$ configuration. Once the initial transitions are overcome, the the spacing between the droplets are observed constant. However, space-time evolution for the $BR=0.75$ \ref{fig:BR_location}b, the film evolution shows significantly different from the behavior shown for $L_p=30D$. Here, the droplet spacing has increased to $\lambda_b=5.8$mm from $\lambda_b=7.7$mm. The evolution shows a Rayleigh-Plateau regime with increased bead spacing compared to figure \ref{fig:BR_influence}c where droplet merging is observed. The dominating waves that appeared as film evolution takes place shows significantly altered behavior based on the location of the bead. The time evolution of the flow rate at downstream locations from the healing is shown in the figure \ref{fig:BR_influence}c. The flow rate variation also shows a periodic variation with  $8$Hz dominating frequency and other harmonic frequencies appearing at two locations as shown in \ref{fig:BR_location}d. This illustrates that location of the blockage will alter the dominating waves that appear on the film growth. To extend the understanding of the bead influence on the film evolution, figure \ref{fig:granular_chain}a,b is shown for two bead and multiple bead configuration. Initially, The spatio-temporal map of the two bead fibre is shown in the figure \ref{fig:granular_chain}a. Here, the beads are positioned in such a way that the droplet have sufficient time to evolve and travel down the second bead on the fibre. Once the droplet crossed second bead, the number of droplet merging events have increased along with the coarsening effect of the droplets. This leads to formation of residual film between two droplets. As the speed of the droplets reduces, the residual film creates ripples due to the Rayleigh-Plateau at the downstream of the second bead. This observation corroborated with the experimental conditions for $25$mm bead spacing shown in the figure \ref{fig:multibead_25mm_spacing}. We have considered multiple bead which represents a granular chain with $BR=0.75$. The bead spacing is chosen below the droplet spacing $\lambda_b$. This configuration lead to coarsening effect of the droplet spacing as shown in the figure \ref{fig:granular_chain}b. The film evolution on the granular chain has significantly altered similar to the experimental observations shown in the figure \ref{fig:multibead_7mm_spacing_exp}. The continuous curvature of the fibre prevented the film from stabilising to form a droplet when more than one bead was closely placed. This finding demonstrates the intricate interplay between bead positioning, spacing and droplet spacing $\lambda_b$ between the droplet.    

\section{Conclusions}
The current findings show that fibre morphologies, specifically single and granular chains with torus beads, have a significant impact on the evolution of liquid films. The analysis and observations are arraived based on the experimental and numerical methodologies. The non-dimensional parameter 'Bead Ratio' ($BR$), which represents the ratio of bead diameter to film height, reveals the selection mechanism that leads to the formation of dominating waves. The transition from regularly spaced droplets to droplet coarsening and merging occurs when both the $BR$ and its distance from the nozzle exceed critical values. Droplet merging mechanisms influenced by the $BR$ for a single bead include the formation of a downstream healing length and its oscillating behaviour.  The presence of beads within the healing length causes behaviour alters the droplet spacing by coarsening but has not altered the Rayleigh-Plateau regime, whereas granular chains significantly alter the film's evolution due to the bead distance's prevention of droplet formation. These findings emphasise the significance of fibre morphology in film development, providing valuable insights for practical applications and contributing to the understanding of fluid dynamics in complex fibre morphology. For many practical applications, it is of interest to consider the three dimensional effects that arise due to fibre morphology, the numerical procedure can readily be extended to three dimension to study the asymmetric instability.

 \textbf{Acknowledgements.} Prof. Gaurav Tomar is gratefully acknowledged for his effort in preparing the manuscript and Prof. Tejas G Murthy for his valuable suggestions.
\bibliographystyle{jfm}
 
\bibliography{jfm}


\end{document}